\documentclass[prb,
                    a4,
                    preprint,
                    showpacs,amsmath,amssymb
                  ] {revtex4-1}

\usepackage{epsfig}
\usepackage{graphicx}
\usepackage{subfigure}

\def\e{\begin{equation}}
\def\f{\end{equation}}

\def\%#1{\mbox{\boldmath $#1$}}
\def\=#1{\overline{\overline #1}}

\def\_#1{{\bf #1}}

\def\.{\cdot}

\def\##1{{\bf#1\mit}}
\def\Re{{\rm Re\mit}}

\def\l#1{\label{eq:#1}}
\def\r#1{(\ref{eq:#1})}
\def\am{\left(\begin{array}{c}}
\def\amm{\left(\begin{array}{cc}}
\def\a{\end{array}\right)}

\begin{document}

\title{Effective electric and magnetic properties of metasurfaces in transition from crystalline to amorphous state}

\author{M. Albooyeh}

\author{D. Morits}

\author{S.A. Tretyakov}

\affiliation{Department of Radio Science and Engineering / SMARAD
Centre of Excellence,\\ Aalto University, P.O. Box 13000, FI-00076
Aalto, Finland}

\date{\today }

\begin{abstract}

In this paper we  theoretically study electromagnetic reflection,
transmission, and scattering properties of periodic and random
arrays of particles which exhibit both electric-mode and
magnetic-mode resonances. We compare the properties of regular and
random grids and explain recently observed dramatic differences in
resonance broadening in the electric and magnetic modes of random
arrays. We show that randomness in the particle positioning
influences equally on the scattering loss from both electric and
magnetic dipoles, however, the observed resonance broadening can be
very different depending on the absorption level in different modes
as well as on the average electrical distance between the particles.
The theory is illustrated by an example of a planar metasurface
composed of cut-wire pairs. We show that in this particular case at
the magnetic resonance the array response is almost not affected by
positioning randomness due to lower frequency and higher absorption
losses in that mode. The developed model allows predictions of
behavior of random grids based on the knowledge of polarizabilities
of single inclusions.


\end{abstract}

\maketitle

\section{Introduction}

Metamaterials are artificial  composite materials which possess
unusual electromagnetic properties not normally found in natural
materials. Electromagnetic properties of nanostructured
metamaterials in the optical range are one of the foci of interest
in modern electromagnetics. Traditionally, metamaterials and
metasurfaces composed of small individual resonant inclusions are
realized as periodical arrays. However, most recently, random or
amorphous metamaterials start to attract attention, see
Refs.~\cite{Helg}--\cite{Chen}. This is due to novel technological
possibilities to manufacture amorphous structures cheaply and on a
large scale, using advanced self-assembly techniques. In addition,
effects of strong spatial dispersion (often undesirable) can be in
some cases suppressed in disordered structures. It is generally
accepted that the electromagnetic properties of both regular and
random arrays of scatterers are quite similar if the distances
between inclusions are electrically small. The main difference in
electromagnetic response comes from scattering on the lattice
inhomogeneities. This apparently results in additional loss in
amorphous metamaterials, and for this reason regular metamaterial
lattices have been the preferred choice if low-loss response is
desired.

However, it appears that in metamaterial structures exhibiting
resonant responses in several modes, the effects due to position
randomness of inclusions are more complicated. In a recent
paper~\cite{Helg} by Helgert et al. reflection and transmission
properties of regular and random (amorphous) planar arrays of
cut-wire particles were studied both numerically and experimentally.
Specially introduced position disorder of individual scatterers
allowed to study the effect of distortion of periodicity on the
electromagnetic response of the array. It was found that position
randomness drastically affects the electromagnetic behavior at the
electric resonance, but makes little impact at the array properties
near the magnetic resonance of the particles. These results were
validated by numerical simulations and confirmed in posterior work
\cite{R2011}.

The authors of paper \cite{Helg} put forward a hypothesis that the
discovered dramatic difference between scattering properties in
electric and magnetic modes is caused by difference in
electromagnetic interactions between particles in different modes.
It was based on an observation that magnetic dipoles as well as
electric quadrupoles do not generate tangential electric fields in
the array plane, and it was assumed that this means that magnetic
scatterers are not interacting with each other, so that the exciting
field acting on a single particle is solely the external
illumination. On the other hand, the electric-dipole scatterers
interact strongly and the exciting field is affected by positional
disorder, which leads to resonance broadening and damping. However,
from the duality principle it is known that in fact magnetic dipole
particles interact via their magnetic fields exactly as strongly as
electric dipoles interact via their electric fields, which means
that the phenomenon discovered in Ref.~\cite{Helg} must have some
other physical reasons.

The goal of this paper is to study the phenomenon of resonance
damping and broadening theoretically and explain the strong
differences in resonance broadening in different resonant modes. To
this end, we analytically study the effect of positional randomness
on electromagnetic behavior of grids of resonant particles which can
exhibit both electric and magnetic resonant responses. We introduce
a simple model, which allows us to analyze the reflective,
transmitting, and absorptive properties of multi-resonant grids,
both in the regular and amorphous states. The theory is confirmed by
numerical simulations using an example of the same metasurface as
that studied in Ref.~\cite{Helg}. The results reveal the mechanisms
of resonance broadening and damping in amorphous structures and
explain the earlier discovered differences in the cases of electric
(symmetric) and magnetic (anti-symmetric) resonances. Understanding
physical phenomena which define the differences between effective
electromagnetic responses of regular and disordered metamaterials is
urgently needed before the emerging amorphous metamaterials can find
applications. Developing analytical models of disordered structures
will allow the design and optimization of future composite materials
with desired performance.

\section{Analytical theory of planar arrays with electrically and magnetically resonant inclusions}

Let us consider an optically dense planar array of optically small
resonant particles excited by normally incident plane waves. We
assume that the distance between the particles in the grid $a$ is
smaller than the wavelength. We are interested in the case when each
particle exhibits both electric and magnetic responses, that is,
both electric and magnetic moments are induced by local electric and
magnetic fields, respectively. We also assume that bi-anisotropic
magnetoelectric coupling is either forbidden due to the particle
symmetry or it is negligible. Many widely-studied infra-red and
optical metamaterial structures like the cut-wire pairs considered
in Ref.~\cite{Helg} belong to this class. In this paper we consider
only electric and magnetic dipole moments of particles, neglecting
quadrupoles and higher-order moments, concentrating on the influence
of array randomness on the reflection and transmission coefficients.
Relative strengths of dipolar and higher-order effects in cut-wire
pairs have been analyzed in Ref.~\cite{Pet}.

Assuming for simplicity that no cross-polarized dipole moments in
the array plane are induced (the particles have the form of discs or
squares, for example) and considering the excitation by normally
incident plane waves, we can write the relations between the induced
electric dipole moment $p$, magnetic dipole moment $m$, and the
incident fields $E_{\rm inc}$ and $H_{\rm inc}$ as scalar relations
\e p=\alpha_{ee}(E_{\rm inc} + \beta_{ee} p), \qquad
m=\alpha_{mm}(H_{\rm inc} + \beta_{mm} m)\l{moments}\f Here
$\alpha_{ee}$ and $\alpha_{mm}$ are the electric and magnetic
polarizabilities of individual inclusions, respectively. Parameters
$\beta_{ee}$ and $\beta_{mm}$ are called \emph{interaction
constants} and they measure contributions of the fields created by
all other particles of the array into the local field $E_{\rm
loc}=E_{\rm inc} + \beta_{ee} p$ exciting each particle (see
e.g.~Ref.~\cite{modeboo}). The interaction constants for electric
and magnetic dipoles are related simply as \e \beta_{mm}={1\over
\eta_0^2}\, \beta_{ee} \f where $\eta_0=\sqrt{\mu_0/\epsilon_0}$ is
the wave impedance of the surrounding space. Fields created by
magnetic dipoles do not contribute to the electric local field
exciting electric dipoles because the tangential component of the
electric field of the magnetic dipole grid equals zero in the array
plane. Likewise, fields scattered by electric dipoles do not excite
magnetically polarizable particles positioned in the same plane.
Most often, both moments are actually induced in the same particles,
but the two modes have resonances at different frequencies.

Next, we calculate the plane-wave electric fields created by the
surface averaged electric current sheet  $J_e=-{i\omega p\over a^2}$
and the magnetic current sheet $J_m=-{i\omega m\over a^2}$ (the
harmonic time dependence assumption is of the form $e^{-i\omega
t}$):
 \e E^{\rm e}_{\rm ref} = -{\eta_0 \over {2}}J_e, \quad H^{\rm
m}_{\rm ref} = -{1 \over {2\eta_0}}J_m \l{EeHm}\f \e E^{\rm m}_{\rm
ref} = -{\eta_0} H^{\rm m}_{\rm ref}, \quad E_{\rm ref} = E^{\rm
e}_{\rm ref}+E^{\rm m}_{\rm ref} \l{EH}\f Here $E^{\rm e}_{\rm ref}$
and $E^{\rm m}_{\rm ref}$ are reflected electric fields created by
the induced electric and magnetic currents $J_e$ and $J_m$,
respectively, and $H^{\rm m}_{\rm ref}$ is the reflected magnetic
field created by the induced magnetic current $J_m$. Solving
\r{moments} for the induced dipole moments in terms of the incident
fields and using \r{EeHm} and \r{EH} we find the reflection and
transmission coefficients in the simple form
%
%
%
%
%
\e R = \frac{E_{\rm ref}}{E_{\rm inc}}= R_e + R_m
 ={i\omega \eta_0 \over 2 a^2}{{1\over {1\over
\alpha_{ee}}-\beta_{ee}}} - {i \omega \over 2 \eta_0 a^2}{{1\over
{1\over \alpha_{mm}}-\beta_{mm}}} \l{R}\f \e T=1+R_e-R_m\l{T}\f Here
we have used the plane-wave relation between the electric and
magnetic incident fields: $H_{\rm inc}=E_{\rm inc}/\eta_0$. The two
partial reflections coefficients $R_e$ and $R_m$ correspond to the
fields created by the induced electric and magnetic currents,
respectively. Since $\beta_{ee}$ has the dimension of $1/(\epsilon_0
a^3)$ and $\beta_{mm}$ has the dimension of $1/(\mu_0 a^3)$, it is
convenient to multiply and divide the reflection coefficients  by
$\epsilon_0 a^3$ or $\mu_0 a^3$.  The result is \e R_e={i k_0 a
\over 2 }{{1\over {\epsilon_0 a^3 \over \alpha_{ee}}- \beta}}
\l{Ten}\f \e R_m=-{ik_0 a \over 2 }{{1\over {\mu_0 a^3\over
\alpha_{mm}}-\beta}} \l{Tmn}\f where
$k_0=\omega\sqrt{\epsilon_0\mu_0}$ is the wave number in the
surrounding space. The normalized dimensionless interaction
constants are the same for both electric and magnetic particles, and
we denote them as $\beta$: \e \beta=\epsilon_0 a^3 \beta_{ee}=\mu_0
a^3 \beta_{mm}\f

Let us assume a simple Lorentz-type resonant response model of
individual particles. This type of resonant response is very common
and approximates very well  the particle response near their
resonances. Let us write down the {\itshape inverse} values of the
normalized polarizabilities to make it easy to discuss the radiation
loss factor: \e {\epsilon_0 a^3 \over \alpha_{ee}} =\left({A_e\over
{\omega_{0e}^2-\omega^2 -i\omega \Gamma_e}}\right)^{-1}-i{k_0^3
a^3\over 6\pi}\l{inv_ee}\f \e {\mu_0 a^3 \over \alpha_{mm}}
=\left({A_m\over {\omega_{0m}^2-\omega^2 -i\omega
\Gamma_m}}\right)^{-1} -i{k_0^3 a^3\over 6\pi}\l{inv_mm}\f Here
$\Gamma_{e,m}$ model the dissipation losses in the particle (in
respective modes), while the last imaginary term is due to the
scattering (re-radiation of power) loss \cite{modeboo}. In case of
regular or ``totally random'' (on the wavelength scale) grids there
is no scattering loss, when the array period is smaller than the
wavelength. In this case spherical-wave scattering from individual
particles is suppressed by interactions between the particles in the
array. Correspondingly, the imaginary parts of the interaction
constants $\beta_{ee}$ and $\beta_{mm}$ contain terms proportional
to $k_0^3$ which compensate the corresponding terms in the inverse
polarizabilities (see e.g. Ref.~\cite{modeboo}): \e
{\beta_{\rm{regular}}= {\rm Re} ( \beta)-i\frac{ k_0^3a^3}{6\pi}+
i\frac{ k_0a}{2}} \l{be_reg}\f The other imaginary term corresponds
to the plane waves created by the surface-averaged currents. In case
of amorphous (on the wavelength scale) arrays particles scatter
individually, and there is no corresponding term in the interaction
constants: \e {\beta_{\rm{amorph}}= {\rm Re}( \beta)+i\frac{
k_0a}{2}} \l{be_amorph}\f In the quasi-static limit $\rm{Re}( \beta)
\approx 0.36$ (see Ref.~\cite{modeboo}).

Next, we substitute these interaction constants and the Lorentz
particle polarizabilities \r{inv_ee} and \r{inv_mm} in the general
formulas for the reflection coefficients \r{Ten} and \r{Tmn}.
For regular or totally random (on the wavelength scale) arrays we
get \e R_{e\,  \rm regular}= i{k_0 a\over 2}{A_e\over \tilde
\omega_{0e}^2-\omega^2 -i\omega \Gamma_e-i\frac{
k_0a}{2}A_e}\f \e R_{m\,  \rm
regular}=-i{k_0 a\over 2}{A_m\over \tilde \omega_{0m}^2-\omega^2
-i\omega \Gamma_m-i\frac{
k_0a}{2}A_m}\f Here $\tilde \omega_0$ denotes the resonant
frequency shifted due to interactions between the particles in the
grid. In the quasi-static approximation for the real part of the
interaction constant $\tilde\omega_{0e,m}^2\approx
\omega_{0e,m}^2-0.36 A_{e,m}$. For amorphous  grids we get \e R_{e\,
\rm amorph}= i{k_0 a\over 2}{A_e\over \tilde \omega_{0e}^2-\omega^2
-i\omega \Gamma_e - {{ik_0^3a^3\over {6\pi}} A_e}-i\frac{
k_0a}{2}A_e}\f \e R_{m\,  \rm amorph}= - i{k_0
a\over 2}{A_m\over \tilde \omega_{0m}^2-\omega^2 -i\omega
\Gamma_m - {{ik_0^3a^3\over {6\pi}} A_m}-i\frac{
k_0a}{2}A_m}\f


Let us consider the case when electric and magnetic resonances occur
at different frequencies. Then in the vicinity of one of the
resonances the non-resonant moment varies weakly with the frequency
and we can find a simple estimation of the resonant curve width (on
the field-strength scale):  \e 2\Delta \omega_{e,m\,  \rm
regular}=\Gamma_{e,m}+\frac{
k_0a}{2}{A_m\over {\tilde \omega_{0e,m}}} \f for regular grids and \e 2\Delta
\omega_{e,m\, \rm amorph}=\Gamma_{e,m}+{k_0^3a^3\over {6\pi}} {A_{e,m}\over
{\tilde \omega_{0e,m}}}+\frac{
k_0a}{2}{A_m\over {\tilde \omega_{0e,m}}}\f for amorphous grids.

We now see that if the condition \e \tilde
\omega_{0e,m}{\Gamma_{e,m}\over A_{e,m}}+\frac{ k_0a}{2}\gg
{k_0^3a^3\over {6\pi}}\l{condition}\f is satisfied, near the
corresponding resonant frequency $\tilde \omega_{0e,m}$ the effect
of inclusion position randomness is negligible, and the response of
regular and amorphous structures is nearly the same. Physically,
this condition means that absorption (the first member of the
left-hand side) and coherent plane-wave reflection (the second
member on the left) dominate over scattering (the right-hand side
term). The above relation shows that this is the case of high
dissipative losses, low resonance strength, and small electrical
size of the unit cell. Note that for the case of negligible
absorption, this condition simply tells that scattering loss is
negligible in random arrays if the distance between particles is
optically very small ($k_0^2a^2\ll 3\pi$).

From the above results we can conclude that the effect of strong
widening of the resonant curve of the electric-dipole mode and
hardly any effect of array randomness on the magnetic mode
discovered in Ref.~\cite{Helg} can be due to two reasons:
\begin{enumerate}
\item At the frequency of the magnetic resonance the grid is
practically homogeneous on the wavelength scale (``totally
random''). Then the scattering term cancels out just like for
periodical grids, and there is no difference in the resonant
curve widths for regular and amorphous layers.

\item At the magnetic resonance the particles are considerably more
lossy and weaker excited than at the electric resonance, that
is, \r{condition} is satisfied near the magnetic resonance but
not satisfied near the electric-mode resonance.
\end{enumerate}

\section{Example: Arrays of cut-wire pairs}

\begin{figure}[h!]
\centering
\includegraphics[width=0.45\textwidth]{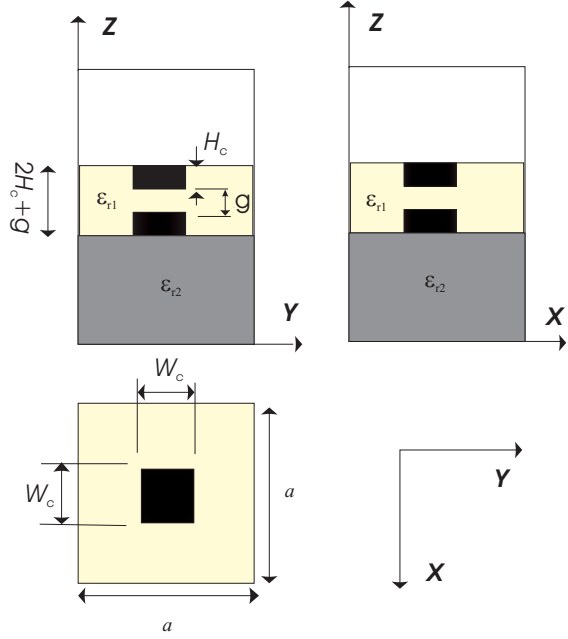}
\caption{(Color online) Geometry of the unit cell of the cut-wire array.} \label{figmodel}
\end{figure}

As an example we consider  the cut-wire pair structure which was
studied in Refs.~\cite{Helg, Shal, Dol}. A unit cell of the infinite
regular array is depicted in Fig.~\ref{figmodel}. The dimensions are
the same as in Ref.~\cite{Helg}. The square lattice has the period
of $a=512$~nm along two transverse directions. The width of the
cut-wire pairs in both lateral directions is $W_c=180$~nm. The
height of each gold pair is $H_c=30$~nm. The gap between the two
elements in each pair equals $g=45$~nm and it is filled with a
material with the relative permittivity equal to
$\epsilon_{r1}=1.72$. The structure is placed on top of a substrate
with the permittivity of $\epsilon_{r2}=1.5$. The permittivity of
gold is taken from Ref.~\cite{John}.

\begin{figure}[h!]
\centering
\subfigure[]{
\includegraphics[width=.45\textwidth]{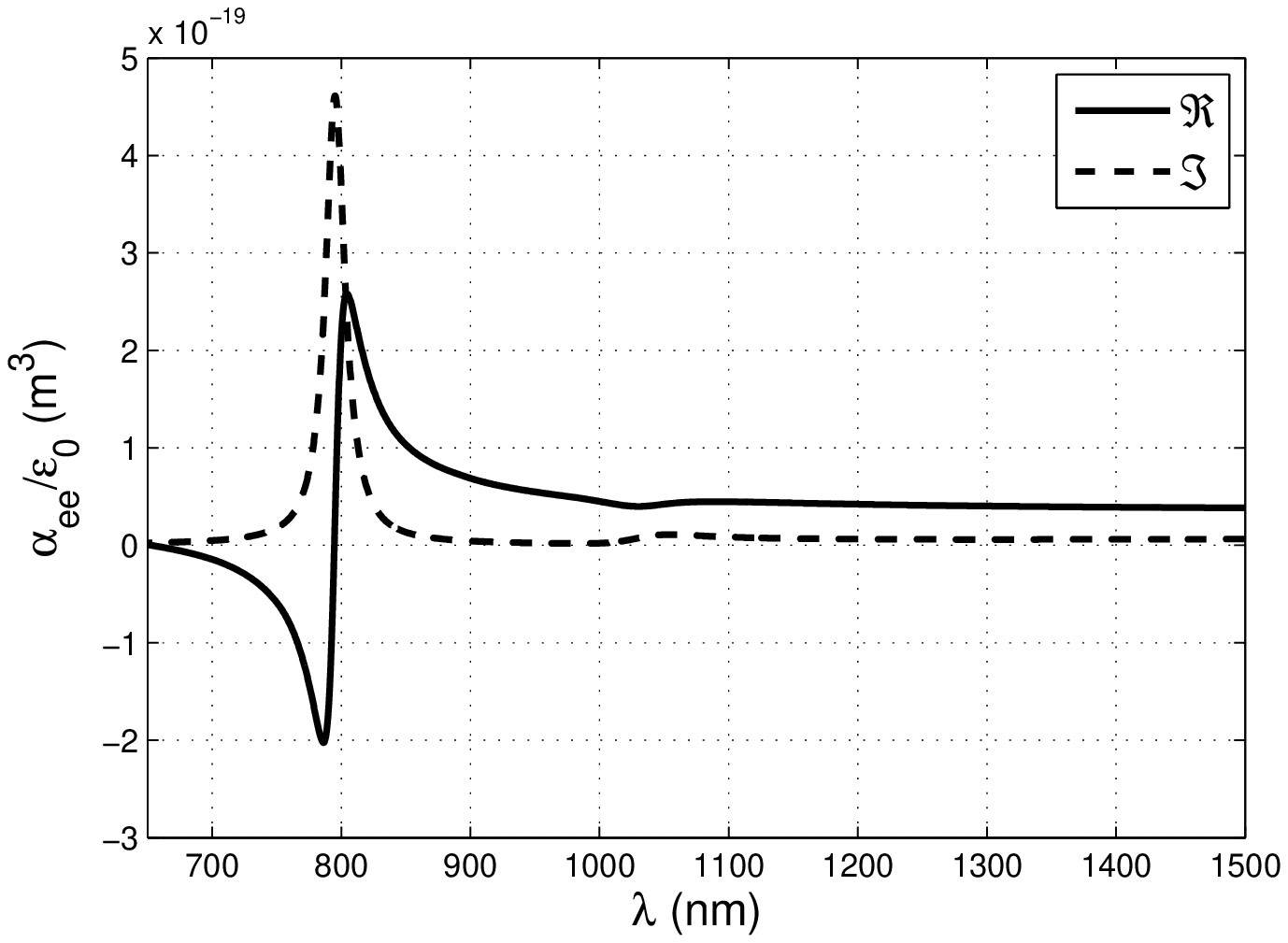}
\label{figalphaE}
}
\subfigure[]{
\includegraphics[width=.45\textwidth]{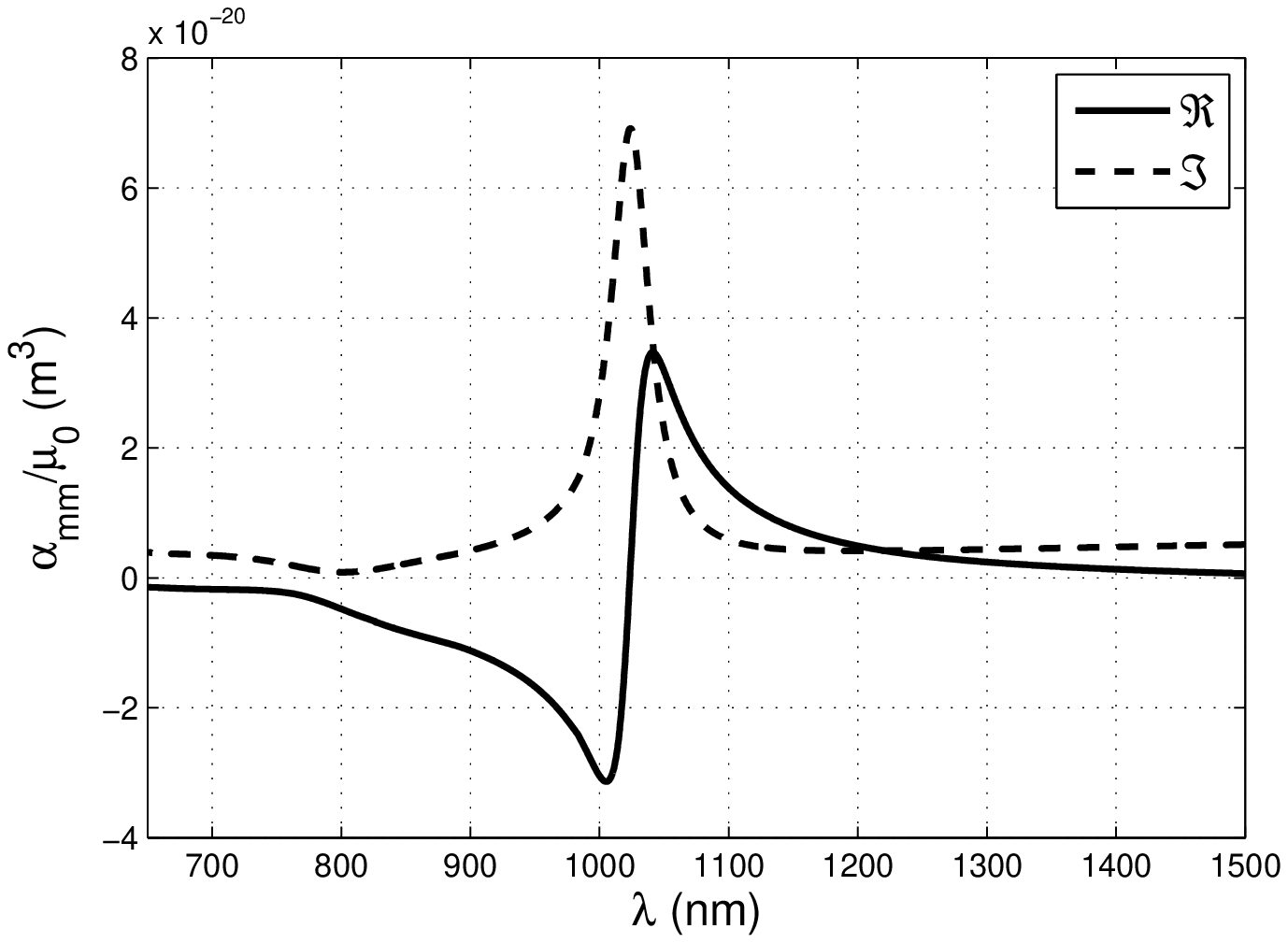}
\label{figalphaM}
}
\caption[Optional caption for list of figures]{\subref{figalphaE}: Electric polarizability of a single cut-wire pair
}
and {\subref{figalphaM}: Magnetic polarizability of the same particle
}
\label{figalpha}
\end{figure}

\begin{figure}[h]
\centering
\subfigure[]{
\includegraphics[width=.45\textwidth]{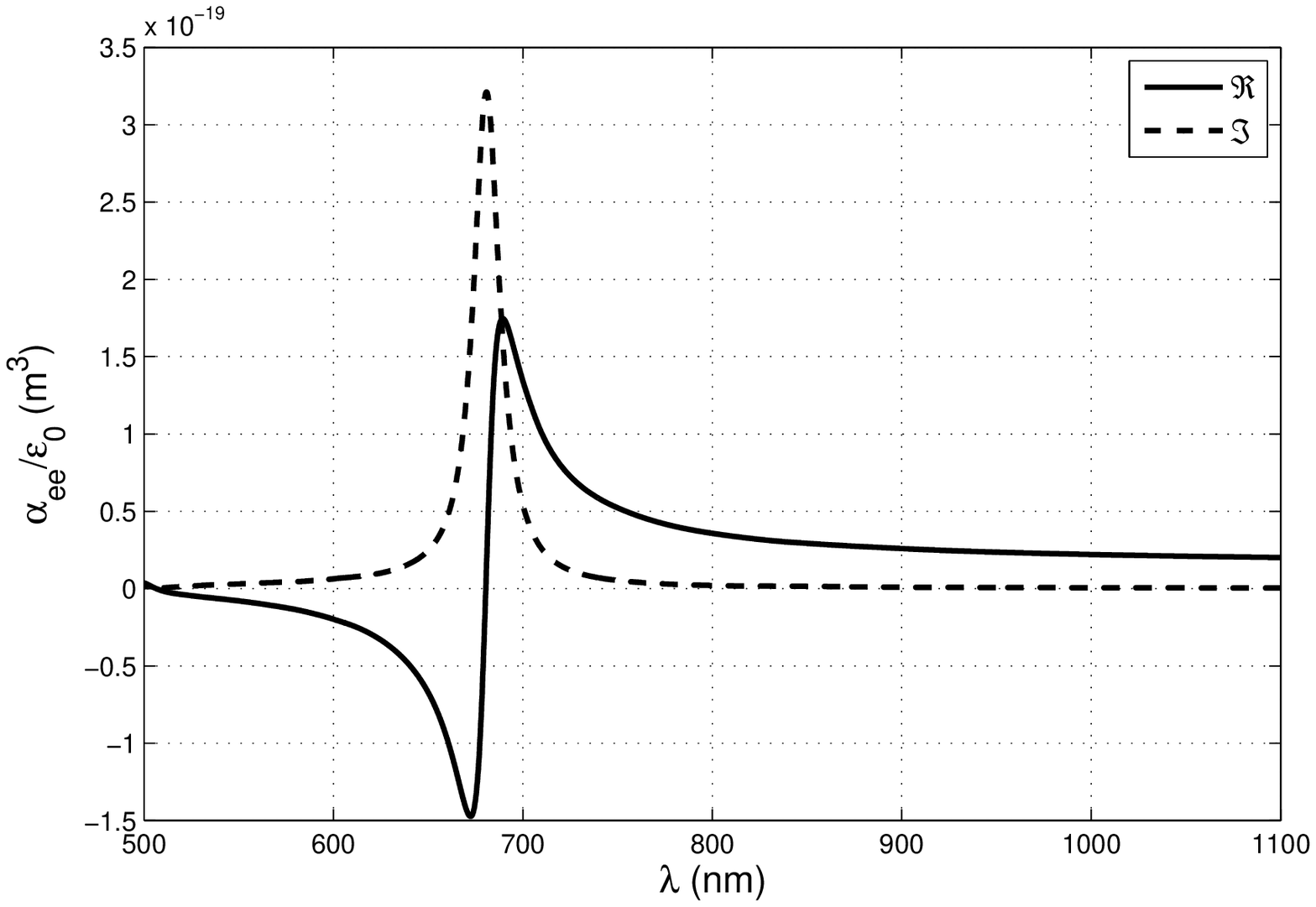}
\label{figalphaEfree}
}
\subfigure[]{
\includegraphics[width=.45\textwidth]{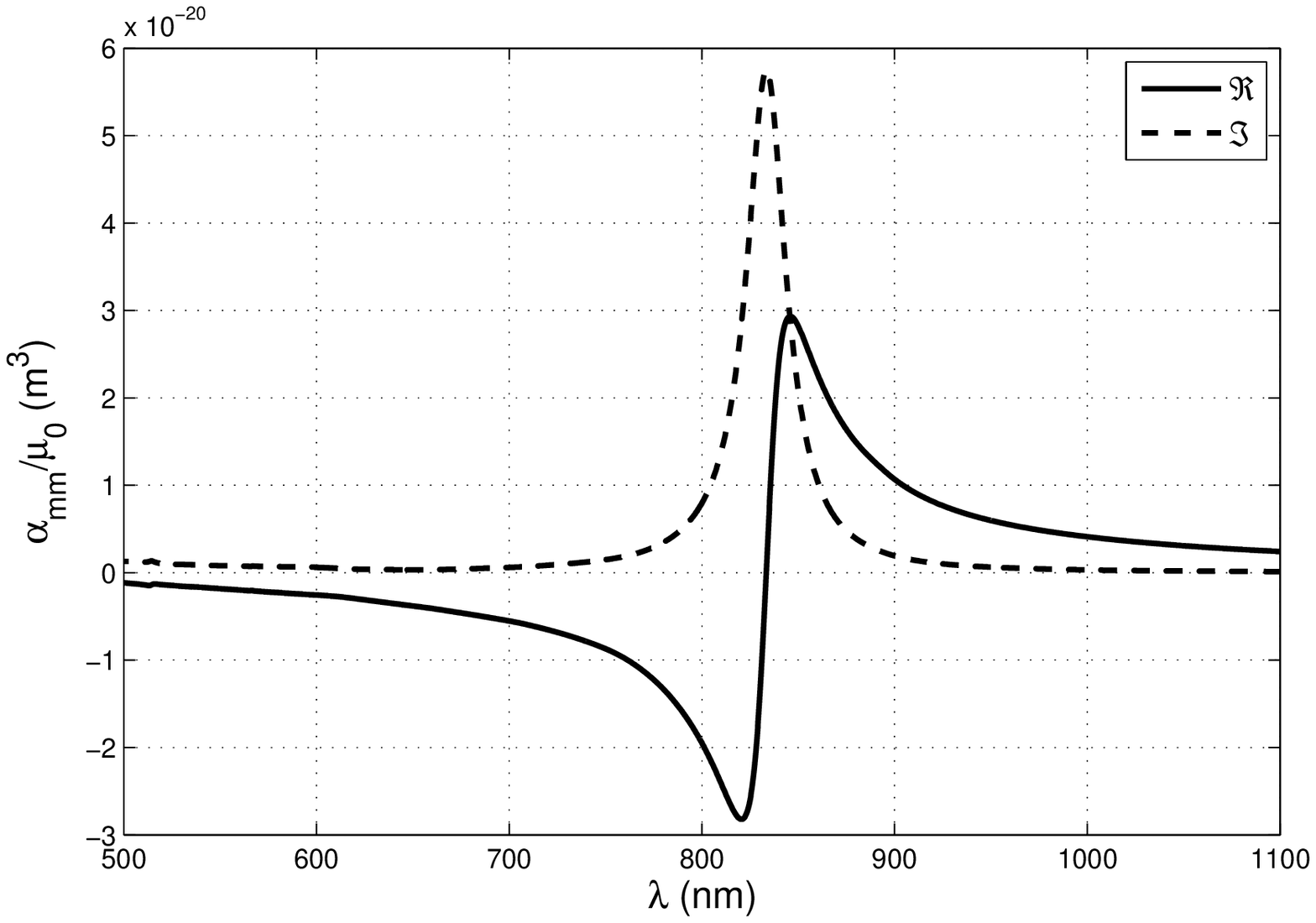}
\label{figalphaMfree}
}
\caption[Optional caption for list of figures]{\subref{figalphaE}: Electric polarizability of a single cut-wire pair in free space and
\subref{figalphaM}:
Magnetic polarizability of the same particle}
\label{figalphaFree}
\end{figure}

\begin{figure*}
\begin{minipage}[h]{0.49\linewidth}
\center
\epsfig{file=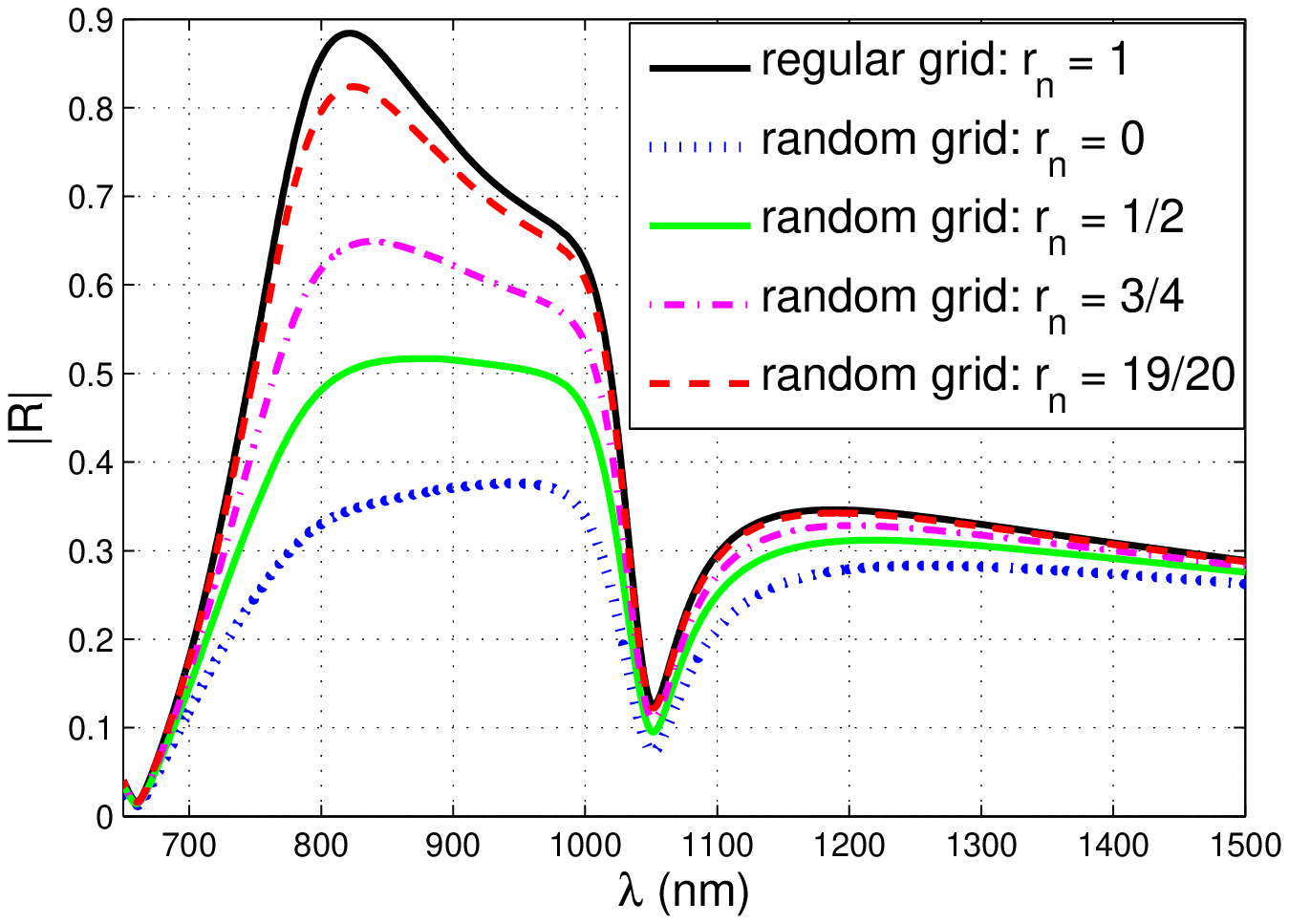,width=1\linewidth, height=0.7\linewidth}
\end{minipage}
\hfill
\begin{minipage}[h]{0.49\linewidth}
\center
\epsfig{file=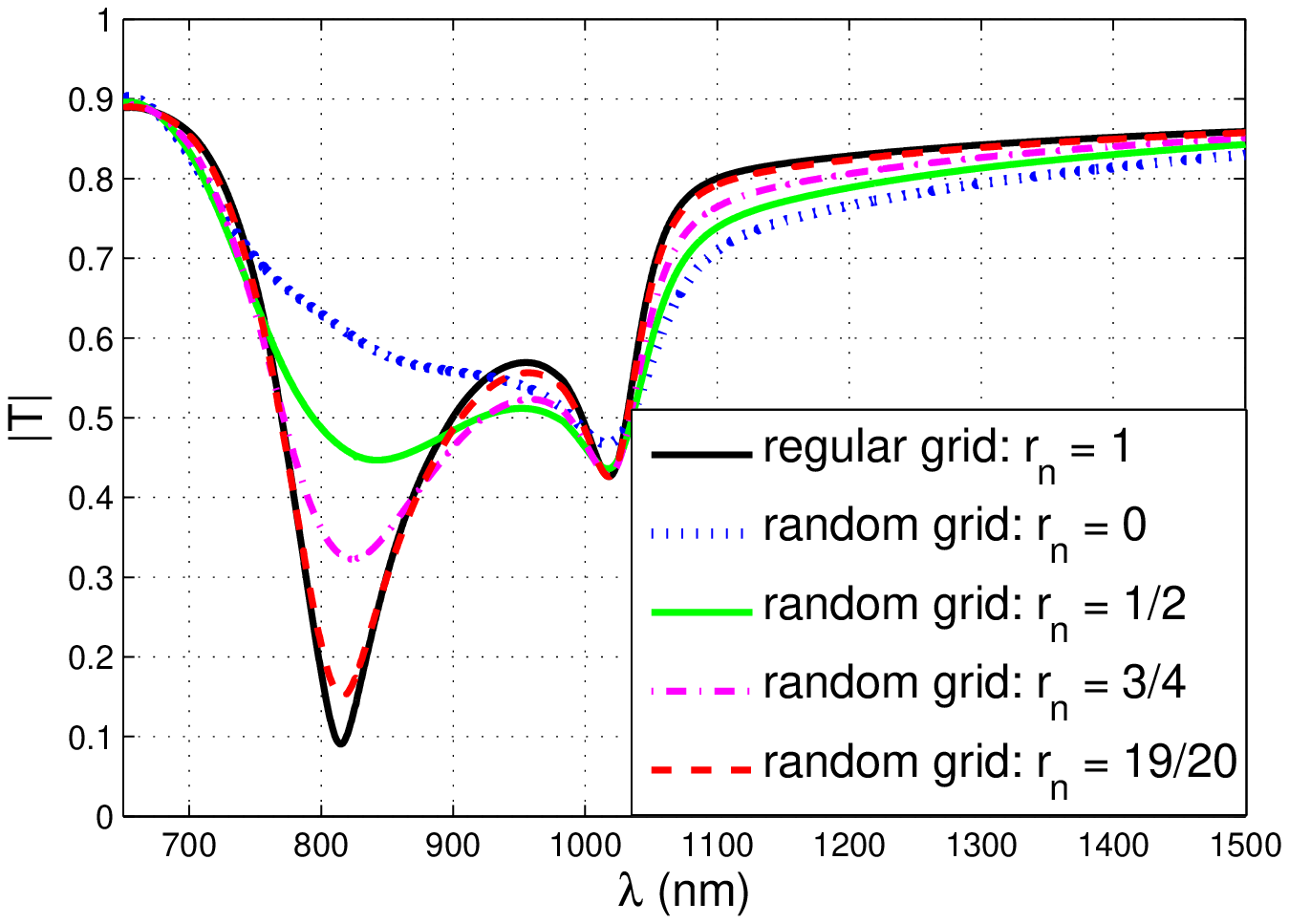,width=1\linewidth, height=0.7\linewidth}
\end{minipage}
\begin{minipage}[h]{1\linewidth}
\begin{tabular}{p{0.49\linewidth}p{0.49\linewidth}}
\centering a) & \centering b)
\end{tabular}
\end{minipage}
\vfill
\begin{minipage}[h]{0.49\linewidth}
\center
\epsfig{file=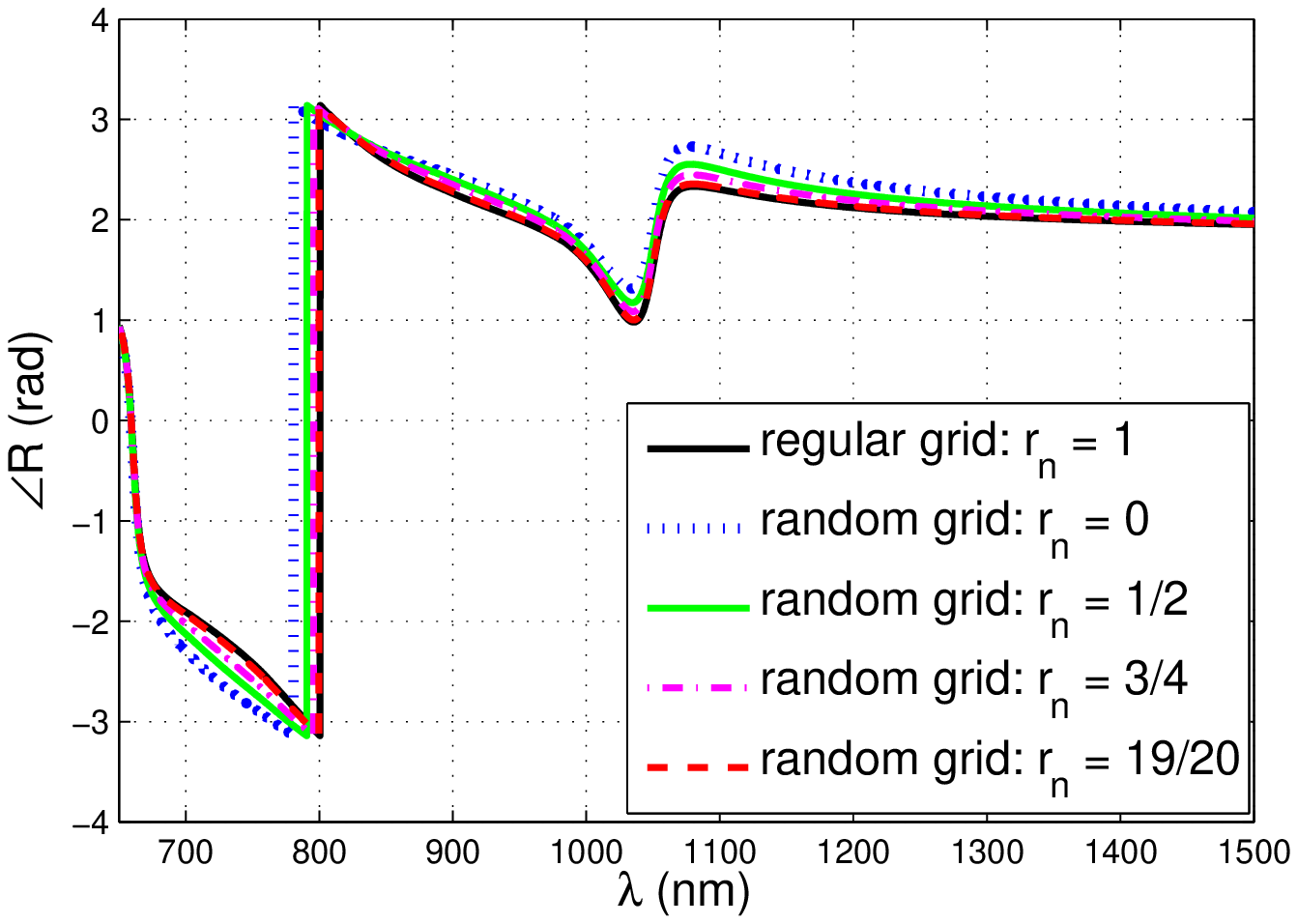,width=1\linewidth, height=0.7\linewidth}
\end{minipage}
\hfill
\begin{minipage}[h]{0.49\linewidth}
\center
\epsfig{file=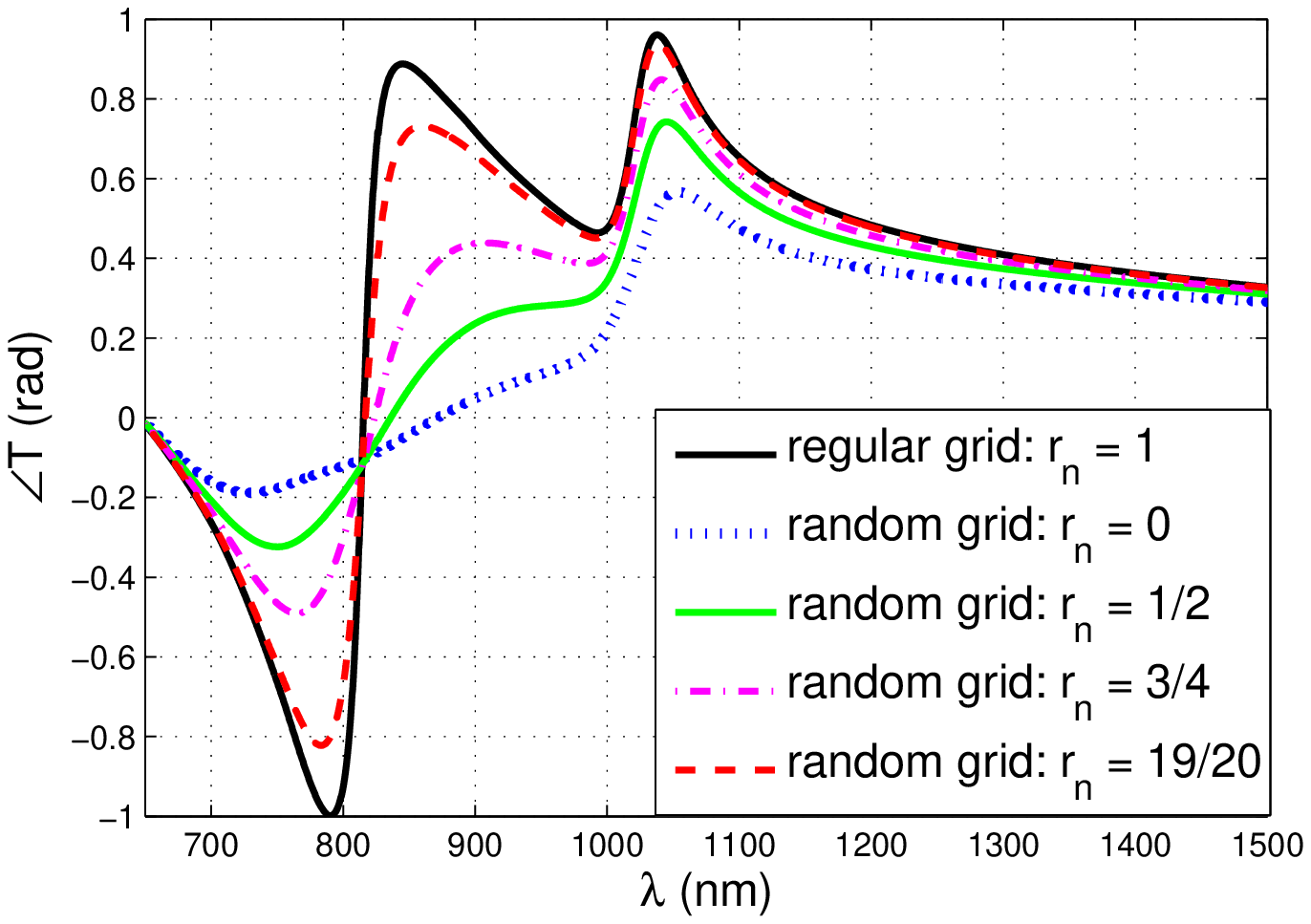,width=1\linewidth, height=0.7\linewidth}
\end{minipage}
\begin{minipage}[h]{1\linewidth}
\begin{tabular}{p{0.49\linewidth}p{0.49\linewidth}}
\centering c) & \centering d)
\end{tabular}
\end{minipage}
\caption{(Color online) (a) -- Amplitude of the reflection coefficient; (b) -- Amplitude of the transmission coefficient; (c) -- Phase of the reflection coefficient; and (d) -- Phase of the transmission coefficient for grids with different randomness levels$(r_n)$} \label{figRTcomp}
\end{figure*}


\begin{figure}[t!]
\centering
\includegraphics[width=0.5\textwidth]{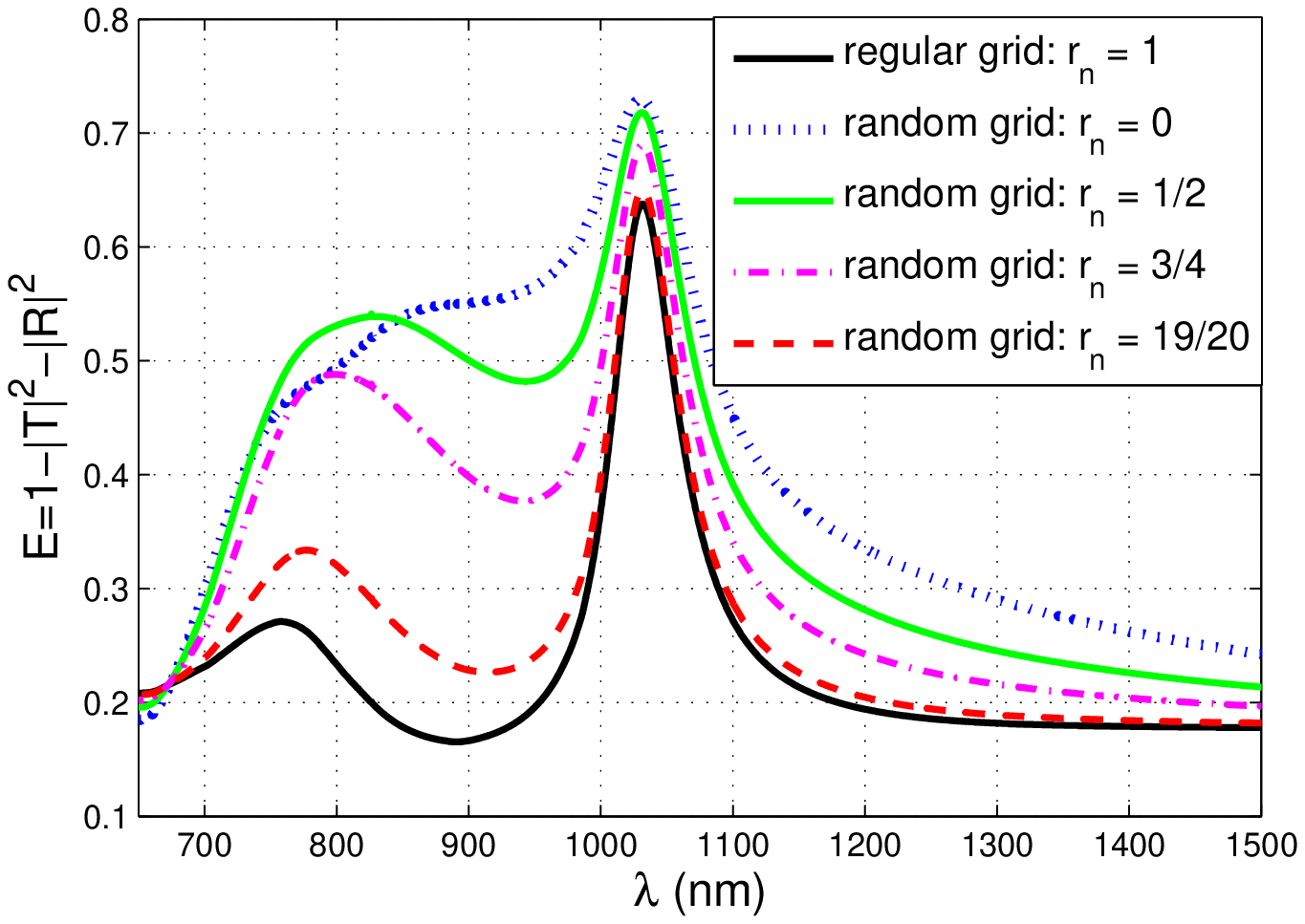}
\caption{{(Color online)}~Absorption in the  array in transition from regular to amorphous states} \label{figEcomp}
\end{figure}

First we calculate the reflection and transmission coefficients for
the regular array using the full-wave numerical simulator Ansoft
HFSS. Substituting the numerical data for reflection and
transmission coefficients in \r{Ten} and \r{Tmn} and using the
quasi-static approximation for the real part of the interaction
constants $\beta\approx 0.36$, we extract the polarizabilities of
individual inclusions. The results are shown in Fig.~\ref{figalpha},
and it is apparent that the array has electric and magnetic
resonances in different frequency regions, as expected. In fact the
array is weakly bi-anisotropic due to the presence of the substrate
(omega-type magnetoelectric coupling~\cite{biama,Mo}), which has
been neglected in the theory and in the parameter extraction. We
have checked that this approximation is valid by repeating the
simulations and parameter extraction for the same array in free
space. The results are presented in Fig.~\ref{figalphaFree} and they
show that this simplifying assumption is reasonable: The substrate
effect is quite small. Behavior of the extracted electric and
magnetic polarizabilities is very close to the canonical
Lorentz-type resonant response.

Scattering losses which appear in transition from regular to
amorphous grids we model by the randomness parameter
$0~\leq~{r_n}~\leq~1$, where unity corresponds to the case where the
scattering loss is completely compensated {\em(regular array)} and
$r_n=0$ means that the scattering loss is not compensated at all
{\em (amorphous array, each inclusion scatters individually)}.
Transition from regular to amorphous state we model modifying the
interaction constant \r{be_reg} as follows:
\e \beta={{\Re\{\beta\}}-{r_n}{i k_0^3a^3\over {6\pi}}+{i k_0a\over
{2}}} \l{betaE}\f
which corresponds to a continuous transition from \r{be_reg} to
\r{be_amorph} with $r_n$ changing from unity to zero. It should be
noted that for simplicity the randomness factor $r_n$ is assumed to
be the same for both electric and magnetic interaction constants.
Due to differences in resonant frequencies, this means that the same
value of $r_n$ may correspond to somewhat different degrees of
geometrical randomness of particle positions.

Next we investigate how the reflection, transmission, and extinction
change in transition from regular to amorphous states, using the
analytical formulas \r{Ten} and \r{Tmn} with the extracted values of
the polarizabilities and the interaction constant \r{betaE}.
Figs.~\ref{figRTcomp} and \ref{figEcomp} show the randomness
effects. One can see that the developed simple model gives very good
agreement with the experimental and numerical data from
Ref.~\cite{Helg}. Electrical response of the grid is strongly
influenced by randomness, while close to the magnetic resonance
there is almost no dependence on randomness. The reason for this
phenomenon is the difference in the ratio of the absorption and
scattering losses. Fig.~\ref{figEcomp} shows that in the periodical
case the absorption is much stronger at the magnetic resonance than
at  the electric one. Higher losses are mainly due to larger
imaginary part of gold permittivity, which is more than two times
higher at the magnetic resonance: $\rm{Im}(\epsilon)_{1022 \,
\rm{nm}} \approx 3.2$, $\rm{Im}(\epsilon)_{797\, \rm{nm}} \approx
1.5$.

In addition, for the case of the grid in free space
(Fig.~\ref{figalphaFree}) we have fitted the numerically extracted
polarizability curves to the Lorentz model \r{inv_ee} and \r{inv_mm}
and extracted parameters $\Gamma_{e,m}$ and $A_{e,m}$. This allowed
us to find the values in inequality \r{condition}. At the resonant
frequency of the electric polarizability we find that the left-hand
side equals $0.4+2.3$ while the right-hand side equals $5.6$.
Scattering effects are clearly dominating and position randomness
changes the array response quite significantly. At the resonant
frequency of the magnetic polarizability the left-hand side reads
$2.3+2$, while the right-hand side equals $3$. In this case the
terms are of the same order and the randomness effect is much
weaker. Note that condition \r{condition} is a simple approximation
which assumes that the two resonances are sharp and well separated.
In this particular example, in the frequency region of the magnetic
resonance the electric dipoles in fact give a significant
contribution to the total absorption and coherent reflection.


\section{Conclusions}

In this paper  we have developed a simple model which explains the
electromagnetic effects in transition from regular to random states
of resonant particle arrays. We have derived a general condition
under which randomizing particle positions gives only negligible
effects on the reflection and transmission coefficients and
explained the earlier discovered dramatic differences in resonance
damping for electric and magnetic modes of particles. We have also
shown that the physical phenomena leading to the resonance damping
in amorphous structures are the same for electrically or
magnetically polarizable particles. The widening of the resonances
takes place due to additional scattering losses, which are
compensated in the case of electrically dense periodical grids.

Studying transition to the amorphous state for a particular example
of cut-wire pairs we have found that the reason for the much weaker
resonance widening and damping in the magnetic mode is strong
absorption in that frequency range. When scattering losses are much
smaller than the dissipation losses, they make little impact on the
total extinction. On the contrary, at the higher-frequency electric
resonance scattering losses are stronger than the dissipation ones,
which leads to strong resonance damping and distortion in the random
case. In other situations, different transition effects in different
resonant modes can also be caused by differences in the electrical
size of the unit cell. In the considered example, the array period
is comparable with the wavelength, thus, even for geometrically
random positions of the particles with respect to the cell centers,
the array cannot be made homogeneous on the wavelength scale.

Our findings can have important implications in understanding the
physical differences in electromagnetic responses of regular and
amorphous structures, in design of various metamaterial structures
for such applications as subwavelength imaging, control of thermal
radiation, microwave, terahertz and optical absorbers, and others.
Using the developed model it is possible to predict and engineer the
effects of randomness, relaxing conventional requirements on strong
periodicity and make use of inexpensive self-assembly techniques in
production of metamaterials.

\subsection*{Acknowledgements}

This study has been done as a student research project within the
Aalto University course on analytical modeling in applied
electromagnetics. One of the authors (ST) wants to acknowledge
enlightening discussions with C. Rockstuhl within the frame of the
EU-funded FP7 project NANOGOLD.





\end{document}